\documentclass[sn-mathphys,Numbered]{sn-jnl}

\usepackage{graphicx}%
\usepackage{multirow}%
\usepackage{amsmath,amssymb,amsfonts}%
\usepackage{amsthm}%
\usepackage{mathrsfs}%
\usepackage[title]{appendix}%
\usepackage{xcolor}%
\usepackage{textcomp}%
\usepackage{manyfoot}%
\usepackage{booktabs}%
\usepackage{algorithm}%
\usepackage{algorithmicx}%
\usepackage{algpseudocode}%
\usepackage{listings}%

\begin{document}

\title[Article Title]{Are you sure? Modelling Drivers' Confidence Judgments in Left-Turn Gap Acceptance Decisions}

\author*{\fnm{Floor} \sur{Bontje}}\email{f.bontje1@uu.nl}
\author{\fnm{Arkady} \sur{Zgonnikov}}\email{a.zgonnikov@tudelft.nl}

\affil{\orgdiv{Department of Cognitive Robotics, Faculty of Mechanical Engineering}, \orgname{Delft University of Technology}, \orgaddress{\street{Mekelweg 2}, \city{Delft}, \postcode{2628 CD}, \country{The Netherlands}}}

\abstract{When a person makes a decision, it is automatically accompanied by a subjective probability judgment of the decision being correct, in other words, a confidence judgment. A better understanding of the mechanisms responsible for these confidence judgments could provide novel insights into human behavior. However, so far confidence judgments have been mostly studied in simplistic laboratory tasks while little is known about confidence in naturalistic dynamic tasks such as driving. 
In this study, we made a first attempt of connecting fundamental research on confidence with naturalistic driver behavior. We investigated the confidence of drivers in left-turn gap acceptance decisions in a driver simulator experiment (N=17). We found that confidence in these decisions depends on the size of the gap to the oncoming vehicle. Specifically, confidence increased with the gap size for trials in which the gap was accepted, and decreased with the gap size for rejected gaps. Similarly to more basic tasks, confidence was negatively related to the response times and correlated with action dynamics during decision execution. Finally, we found that confidence judgments can be captured with an extended dynamic drift-diffusion model. In the model, the drift rate of the evidence accumulator as well as the decision boundaries are functions of the gap size. Furthermore, we demonstrated that allowing for post-decision evidence accumulation in the model increases its ability to describe confidence judgments in rejected gap decisions. Overall, our study confirmed that principles known from fundamental confidence research extend to confidence judgments in dynamic decisions during a naturalistic task.}


%




\maketitle

\section{Introduction}\label{sec1}
Our decisions are automatically accompanied by a confidence judgment which is one's subjectively estimated probability that their decision is correct~\cite{ fetsch_predicting_2014,yeung_metacognition_2012,kepecs_computational_2012}.
Previous research has demonstrated that the processes responsible for decision making and confidence judgments are closely related to each other~\cite{kepecs_computational_2012,kiani_representation_2009,yeung_metacognition_2012,murphy_neural_2015}. Confidence judgments affect for example the justifications of future decisions and behavior~\cite{folke_explicit_2017}. Moreover, the processes responsible for decision making~\cite{ratcliff_diffusion_2008} and confidence judgments are both dependent on the accumulation of evidence towards or against a decision~\cite{yeung_metacognition_2012,pleskac_two-stage_2010,murphy_neural_2015,fleming_self-evaluation_2017,kobe2020,de_martino_confidence_2013}. 

Earlier experimental studies and confidence models provided many insights into confidence judgments. In particular, two of the most robust findings in confidence research is that confidence in a decision increases with the strength of evidence in favor of the chosen option and that confidence correlates negatively with response time~\cite{rahnev_confidence_2020,kiani_choice_2014,yeung_metacognition_2012, fleming_metacognition_2023}. However, these findings have been replicated in traditional laboratory setups with simplistic preferential~\cite{de_martino_confidence_2013,brus_sources_2021}, inferential~\cite{handel_individual_2020,liberman_local_2004,chua_building_2015}, or perceptual tasks~\cite{rouault_forming_2019, kiani_choice_2014}. At the same time, it is unclear if the empirical findings and computational models of confidence in simple tasks generalize to naturalistic, real-life decisions which typically involve complex task structure and dynamically changing perceptual information~\cite{fleming_metacognition_2023}. 

One domain in which human decisions are inherently dynamic is traffic: human drivers and pedestrians make safety-critical decisions in complex dynamic scenes on a daily basis. Recent research has examined the cognitive mechanisms behind the decision-making processes underlying traffic decisions, revealing that evidence accumulation models can explain decisions and response times observed in a variety of tasks such as pedestrian crossing~\cite{pekkanen_variable-drift_2022, markkula_explaining_2023}, left turn maneuvers~\cite{zgonnikov_should_2022, zgonnikov_subtle_2023}, and overtaking~\cite{mohammad_cognitive_2023}. In the context of lane-change decisions, driver \textit{uncertainty} judgments (defined as ``driver’s difficulty to make appropriate decisions of either changing the lane or not in a given lane change situation'') have been shown to depend on kinematics of vehicles in the traffic scene~\cite{yan_classifying_2015,yan_investigating_2023}. Uncertainty judgments have also been investigated in overtaking decisions~\cite{leitner_overtake_2023} and interactions at narrow passages~\cite{miller_time_2022}. However, uncertainty judgments in traffic decisions have so far been analyzed disregarding decision outcomes, obscuring the relationship between these judgments and task parameters. Furthermore, despite the similarity of concepts, the exact relationship between (un)certainty and confidence in decision making is unclear~\cite{pouget_confidence_2016}. This hampers linking recent findings on uncertainty in driving decisions to the fundamental findings and computational models of confidence. 

In summary, confidence judgments in traffic decisions have not been measured yet, let alone modelled. This renders it unclear whether basic mechanisms of confidence extend to complex tasks such as driving. In addition, the lack of understanding of cognitive mechanisms underlying confidence hinders practical applications such as advanced driver assistance systems and automated vehicles, the design of which could potentially utilize the insights into drivers' confidence.

In this study, we made a first attempt of connecting the fundamental research on confidence with studies of naturalistic human driver behavior. We investigated the confidence of human drivers in left-turn gap acceptance decisions at unprotected intersections. We examined the influence of the time-to-arrival (TTA) and the distance gap to the oncoming vehicle 
on the decisions (Research Question 1a, RQ1a) and response times (RQ1b) of drivers. 
Additionally, we assessed how confidence is related to the time-to-arrival and distance gap (RQ2) and how confidence is related to the response time (RQ3). Moreover, we investigated how action dynamics of drivers' behavior after the decision relate to confidence (RQ4). 
Last but not least, we assessed how well confidence judgments of drivers in left-turn gap acceptance decisions can be described by four candidate cognitive models based on extended evidence accumulation models (RQ5). Specifically, we tested models using either one evidence accumulator describing both decisions (drift-diffusion model) or two independent competing evidence accumulators (race model)~\cite{zgonnikov_should_2022,zylberberg_construction_2012,bogacz_physics_2006}. Also, we investigated for each model whether it performs best with or without making use of post-decision evidence accumulation ~\cite{murphy_neural_2015,fleming_self-evaluation_2017}.

\section{Experiment: Methods}\label{sec2}
In this study, we investigated the confidence of drivers in left-turn gap acceptance decisions through a fixed-base driver simulator experiment ($N=17$; $9$ male; $8$ female; mean age $31\pm \textrm{(std)}\, 11$ years). The experiment followed the 2-by-2 within-subjects design. All participants were in possession of a valid driving license. The Human Research Ethic Committee of the TU Delft approved the study and participants signed an informed consent form prior to the experiment. All the data and source code used to collect and analyze the data are publicly available at \url{https://osf.io/tgexp/}.

\subsection{Setup}
For the driving simulation, we used the simulator software Carla~\cite{dosovitskiy_carla_2017} on a Windows-based desktop computer. The driver simulator hardware consisted of a 65-inch screen (Samsung UE65MU7000) and a Logitech G29 steering wheel and pedals set. Participants were seated at a distance of about 1.3 meters from the screen. 
\begin{figure}[ht!]
    \centering
    \includegraphics[clip, trim=0.2cm 0.1cm .2cm 0cm,width = 0.6\linewidth]{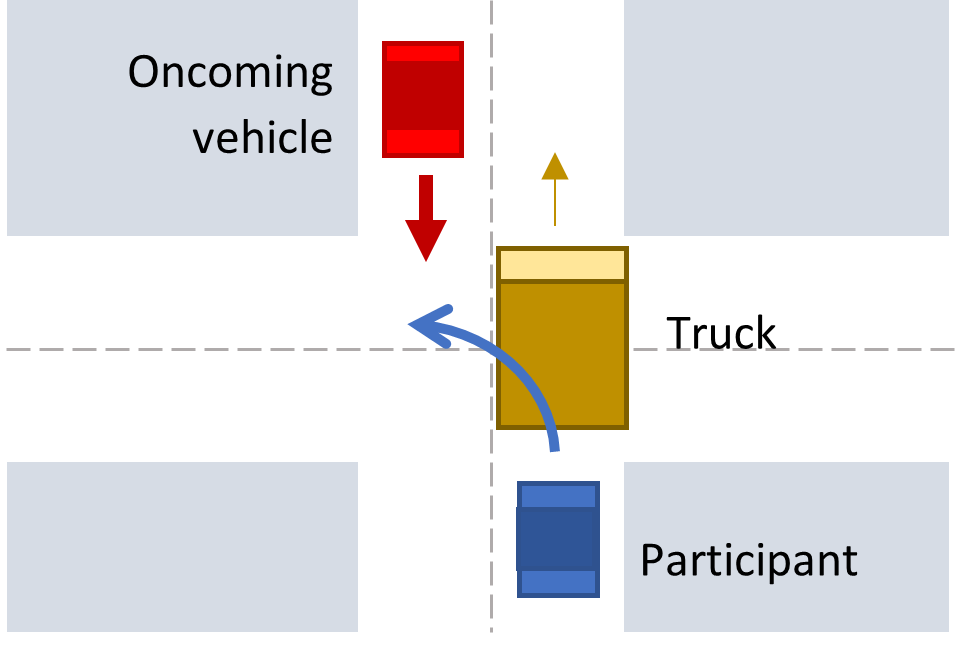}
    \caption{Top-down view of the left-turn gap acceptance scenario. As the truck moved away from the intersection, the oncoming vehicle appeared in participant's field of view.}
    \label{fig:top_view}
\end{figure}
\begin{figure*}
\centering
\includegraphics[width = 1\linewidth]{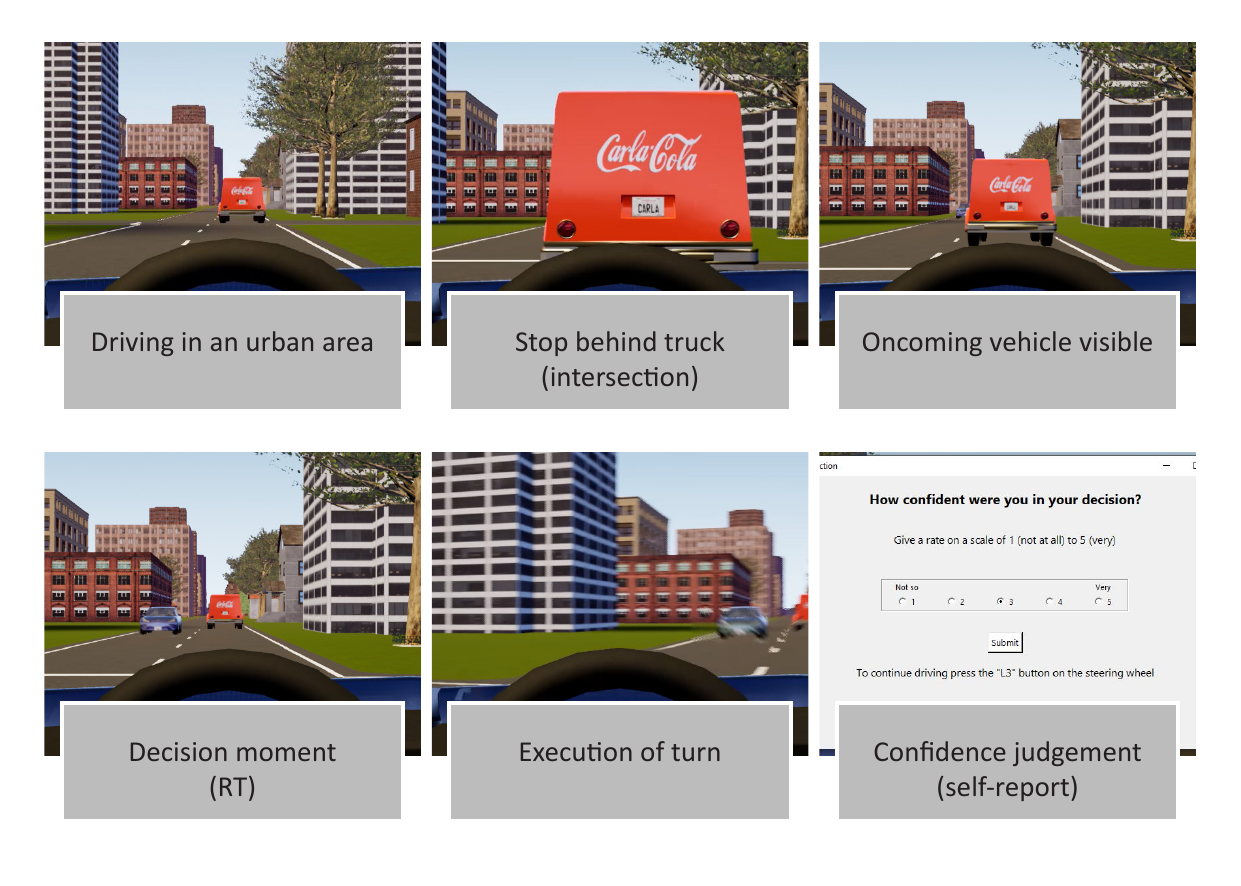}
\caption{Participants' view at different moments throughout a trial: 1) Driving through the urban environment, 2) Stop at an intersection, whereby the front view is blocked by the truck, 3) The truck drives away, revealing the oncoming vehicle (the decision-making process starts), 4) Participant has to decide to ``go'' or ``wait'' by pressing a button on the steering wheel (the decision-making process ends; response time is recorded), 5) Execution of the turn, 6) Self-report on the confidence the participant had in the made decision by answering the question: ``How confident were you in your decision? Give a rate on a scale of 1 to 5''.}
\label{fig:phases_exp}
\end{figure*}

\subsection{Experiment protocol}
The participants were instructed to drive as they normally would, following auditory prompts which navigated them through a virtual urban area. The area consisted of a square grid of blocks (150 by 150 meters each). Before each intersection, at a distance of 120 meters and 30 meters, navigation prompts were provided through the headphones, instructing the participant to drive straight ahead or to make a left or right turn. 

Each participant drove five different randomly generated routes, with each route comprising 20 intersections. At 80\% of the intersections, the participant had to make a left-turn across the path of an oncoming vehicle (Figure~\ref{fig:top_view}). In the remaining intersections, participants were instructed to drive straight ahead (10\%) or to make a right-turn (10\%). Before entering an intersection, the participants needed to stop before a truck that was positioned at the centre of the intersection. The truck was present in order to limit the view of the driver and to make the oncoming car appear in a natural manner. The participant's vehicle, the truck, and the oncoming vehicle were the only vehicles in the scene, i.e. no other vehicles were behind the oncoming vehicle.

To make a left turn, participants had to first make a gap acceptance decision, that is, whether to accept the gap and perform the turn in front of the oncoming vehicle (``go'' decision) or reject the gap and wait until the oncoming vehicle had passed (``wait'' decision) (Figure~\ref{fig:phases_exp}). Participants were asked to press one of the two buttons on the steering wheel as soon as they made the decision, and execute the decision immediately after that. After the participants finished the turn (regardless of whether this was done before or after the oncoming vehicle), they were asked to self-report the level of confidence in their decision. This was the only measurement of confidence we collected.  

During the driving task, we altered the traffic situation by varying the distance gap and the time-to-arrival (TTA) of the oncoming vehicle. The distance gap and time-to-arrival were the two independent variables, taking values of 70 and 90 meters and 5.5 and 6.5 seconds, respectively. On each route, each combination of distance and time-to-arrival was present four times in the random order. The ratio between the initial distance gap and time-to-arrival determined the velocity of the oncoming vehicle, which ranged between 38.77 km/h and 58.91 km/h and remained constant throughout the trial.

To become comfortable with the task, the participants drove at least one practice route (that is, ten intersections) before starting the experiment. We mitigated the fatigue and habituation effects by including a break halfway through the experiment and by offering the participants the opportunity to take more breaks if needed. 

\subsection{Data analysis}
During the experiment, we recorded the positions, velocities, and accelerations of all the vehicles (frequency: 100 Hz). We also recorded the gas throttle input of the participants’ vehicle.

\subsection{Metrics}
The dependent variables in this study were the decision outcome (``go''/ ''wait''), the response time (RT), and the confidence. In addition, we assessed the relationship between the reported confidence and action dynamics by analyzing the velocity of the participants' vehicle and the distance to the centre of the intersection while executing the turn.

\subsubsection {Decision outcome, response time, and confidence}
We instructed participants to report their decision as soon as they decided by pressing one of the two designated buttons on the steering wheel. We defined response time as the time between the moment the oncoming vehicle appeared in participant's field of view and the moment one of the buttons was pressed.

We measured the level of confidence participants had in their decision by presenting them with the question ``How confident were you in your decision? Give a rate on a scale of 1 to 5''. We posed the question after the turn was performed, at a distance of ten meters from the centre of the intersection, in order to minimise the interruption of the driving task. Participants had to provide the confidence judgment before they could resume the driving task.

\subsubsection{Action dynamics}
In order to analyse how the response time and decision confidence are reflected in the subsequent driving behavior, we investigated the action dynamics related to the execution of the turn: the velocity profile and the distance to the centre of the intersection.

The velocity profile during the turn described the absolute velocity of the participants’ vehicle over time. In all analyses of the velocity profile, we used the initial throttle operation moment as the zero-point in time in order to negate the effect of the response time on action dynamics. As the duration of the turn execution varied strongly within and across participants, we only analyzed the velocity profiles until the cutoff point ($t_{\text{cutoff}}$), defined as 0.75-percentile of the turn execution duration (calculated separately for ``go'', $t_{\text{cutoff}}$ = 3.24 seconds, and ``wait'' trials, $t_{\text{cutoff}}$ = 6.11 seconds).

To analyse the relation between confidence and the velocity profile, we used the following trial-level metrics: the maximum absolute velocity value during a trial, the signed average deviation from the individual mean ($\textrm{DM}_\textrm{indiv}$, Eq.~\eqref{eq:DMindiv}), the signed average deviation from the group mean ($\textrm{DM}_\textrm{group}$, Eq.~\eqref{eq:DMgr}), and the root mean square deviation from the individual mean ($\textrm{RMSD}_\textrm{indiv}$, Eq.~\eqref{eq:rmsd}).

Given a velocity vector $y=(y_1, ..., y_{N_t})$ of length $N_t$,
\begin{equation}\label{eq:DMindiv}
    \textrm{DM}_\textrm{indiv} = \frac{\sum_{i=1}^{N_t} (y_i-\mu_\textrm{indiv}^{\textrm{decision}})}{N_t},
\end{equation}
\begin{equation}\label{eq:DMgr}
    \textrm{DM}_\textrm{group} = \frac{\sum_{i=1}^{N_t} (y_i-\mu_\textrm{group}^{\textrm{decision}})}{N_t}.
\end{equation}

\begin{equation}\label{eq:rmsd}
    \textrm{RMSD}_\textrm{indiv} = \sqrt{\frac{\sum_{i=1}^{N_t} (y_i-\mu_\textrm{indiv}^{\textrm{decision}})^{2}}{N_t}},
\end{equation} 

The signed deviations from the mean (Eq.~\eqref{eq:DMindiv} and Eq.~\eqref{eq:DMgr}) and the root mean square deviation (Eq.~\eqref{eq:rmsd}) indicate how the velocity profile of a given trial deviates from the average trial. We computed the deviation from the individual mean (Eq.~\eqref{eq:DMindiv}) and root mean square deviation by comparing the velocity profile of the trial with the individual participant's average velocity profile in all trials with the corresponding decision outcome (``go''/ ''wait''). At the same time, the deviation from the group mean (Eq.~\eqref{eq:DMgr}) quantifies the difference of the velocity profile in a trial to the group-averaged velocity profile calculated over all participants. We included mean deviations from both individual and group means to better distinguish within-participant effects from effects due to individual differences.

The main advantage of using (signed) deviations from the mean ($\textrm{DM}_\textrm{indiv}$ and $\textrm{DM}_\textrm{group}$) is that they provide an indication of the extent and the direction of deviation. These metrics, however, level out positive and negative values of the deviation. The root mean square deviation prevents this, but does not provide an indication of the direction of the deviation, thereby complementing the signed average deviation metrics. 

For the analyses of the distance to the centre of the intersection, we used the same metrics as for the velocity profile (calculated using the same alignment and cutoff time points) with the exception of using the minimum absolute value rather than the maximum value. The latter metric allowed us to capture corner-cutting behavior during the turn execution. 

\subsection{Statistical analysis}
We analyzed the relationships between the dependent and independent variables using mixed-effects models, using random intercepts or slopes per participant (ID) to account for between-participant differences. All statistical analysis were performed in MATLAB R2020a using the generalized linear mixed effects model \texttt{fitglme} and linear mixed effects model \texttt{fitlme} functions. In models including decision as a predictor, ``go'' was used as a reference class. 

We used the following models for decision, response time, and confidence, respectively: 
\begin{equation}\label{eq:lme_pr}
    \textrm{decision} \sim \textrm{distance} + \textrm{TTA} + (1|\textrm{ID})
\end{equation}

\begin{equation}\label{eq:lme_RT}
     \textrm{RT} \sim \textrm{decision}*(\textrm{distance} + \textrm{TTA}) + (\textrm{decision}|\textrm{ID})   
\end{equation}

\begin{equation}\label{eq:lme_conf_rt}
    \textrm{confidence} \sim \textrm{decision}* (\textrm{RT} + \textrm{distance} + \textrm{TTA}) + (\textrm{decision}|\textrm{ID})
\end{equation}

To investigate the relationship between the action dynamics metrics and confidence, we used linear mixed-effects models with the following structure: 
\begin{equation}\label{eq:metric}
    \textrm{metric} \sim \textrm{confidence}*\textrm{decision} + (1|\textrm{ID})
\end{equation}

\subsection{Hypotheses}
\subsubsection{Decision behavior and response times}

We expected to observe a positive relation between the time-to-arrival and distance gap conditions and the probability of ``go'' decision (Hypothesis 1.1, H1.1), and a positive relation between the response time and the time-to-arrival (but not distance~\cite{zgonnikov_should_2022}) for both decisions (H1.2). These hypotheses are based on the previous studies of left-turn and overtaking gap acceptance decisions which found that the probability of accepting the gap as well as the response times are affected by the perceptual information capturing the time-to-arrival and distance gap~\cite{zgonnikov_should_2022,zgonnikov_subtle_2023,sevenster_response_2023}. 

\subsubsection{Effect of time-to-arrival and distance gap on confidence} 

Earlier research has shown that confidence judgments relate to the available perceptual evidence towards a decision, and follows the ``x''-pattern: confidence increases with the strength of the evidence for decisions consistent with that evidence, and decreases with the strength of the evidence if the decision goes against the evidence~\cite{fleming_self-evaluation_2017, yeung_metacognition_2012}. Based on this, we hypothesised that confidence positively relates to the time-to-arrival and the distance gap of the oncoming vehicle in ``go'' decisions, and that it negatively relates to these factors in ``wait'' decisions (H2.1). Intuitively, a larger time-to-arrival and distance gap give the driver more time and space to perform the left-turn in front of the oncoming vehicle, and therefore are hypothesized to result in greater confidence when the gap is accepted. However, if the driver rejects the gap despite its large size, the confidence in that decision is expected to be low.

\subsubsection{Relationship between confidence and response time}
Earlier confidence research has consistently revealed negative relationship between response times and confidence across trials: the longer participants take to make the decision, the lower their confidence is~\cite{pleskac_two-stage_2010}.  
We therefore hypothesized that response time will be negatively related to confidence across trials (H3).

\subsubsection{Relationship between confidence and action dynamics}
Most studies of confidence have so far used paradigms that provide little data beyond decision outcomes and response times. However, more dynamic paradigms capturing continuous trajectories of decision execution are becoming increasingly popular in decision-making research, including tracking mouse cursor~\cite{freeman_doing_2018,schulte-mecklenbeck_process-tracing_2017}, reaching~\cite{song_hidden_2009, wispinski_models_2020}, and full body movements~\cite{zgonnikov_beyond_2019}. Such paradigms have begun to unveil the link between confidence and action dynamics as expressed in hand movements~\cite{van_den_berg_common_2016,dotan_-line_2018,dotan_track_2019}. Following these studies, we expected that confidence in our paradigm will be related to the action dynamics of turn execution. In particular, we hypothesised:

\begin{itemize}
    \item in ``go'' decisions, a positive relation between confidence on the one hand and the maximum velocity (H4.1a) and average deviation from the mean velocity (H4.2a) on the other hand. Furthermore, we hypothesised a negative relation between confidence and minimum distance to the center of the intersection (H4.1b) and average deviation from the mean distance (H4.2b). Here, we expected the participants to \textit{drive more confidently} (e.g., with higher speed and following a straighter path) when executing higher-confidence ``go'' decisions.
    \footnote{When testing H4.2a and H4.2b, we used average deviation from individual mean values (Equation~\ref{eq:DMindiv}) as the main metric of interest; in addition, we analyzed deviation from group mean values (Equation~\ref{eq:DMgr}) to highlight potential effects due to individual differences between participants.}

    \item a negative relation between confidence on the one hand and the root mean square deviation of the velocity profile (H4.3a) and distance to the centre of the intersection (H4.3b) on the other hand. In other words, we expected participants to deviate stronger from their typical behavior in trials where they reported lower confidence.
    
    \item a weaker relation between confidence and the action dynamics measures in ``wait'' decisions compared to ``go'' decisions (H4.4). The rationale for this is that execution of the turn immediately follows the ``go'' decisions but is separated from ``wait'' decisions by several seconds of inaction.
  
\end{itemize}

\section{Experiment: Results}\label{sec3}
\subsection{Decision behavior and response times}

\begin{figure}[ht!]
    \centering
    \includegraphics[clip, trim=1.4cm 9cm 2cm 9.5cm, width = 0.7\linewidth]{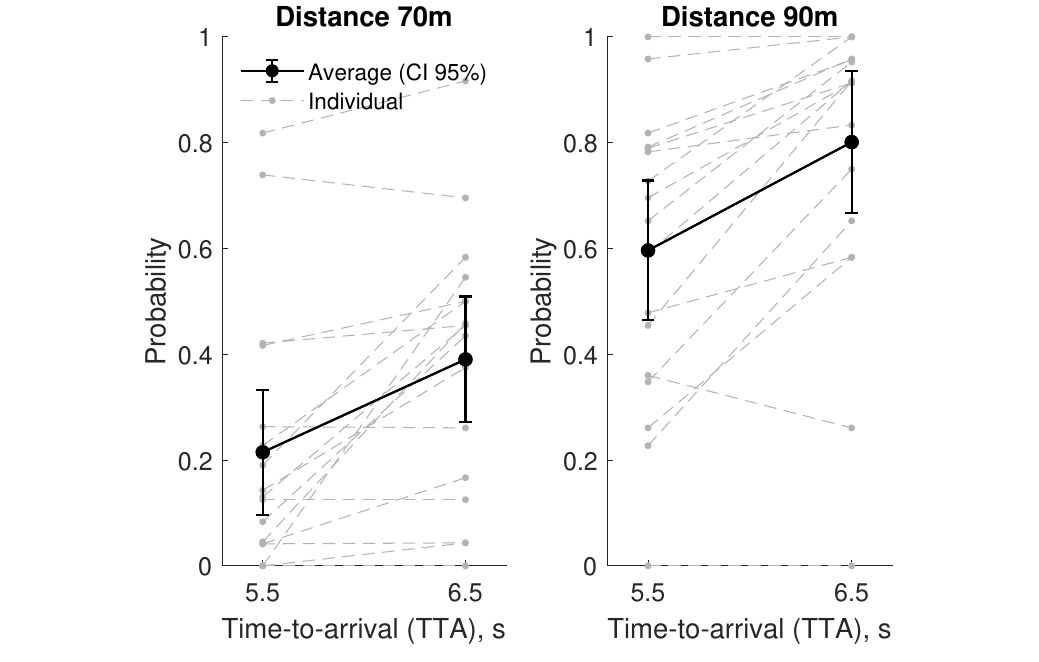}
    \caption{Group-averaged (solid lines) and individual participants' (dashed lines) probabilities of ``go'' decision as a function of the time-to-arrival and distance gap to the oncoming vehicle. Error bars represent 95\% CI over $n=17$ individual participants' mean values. }
    \label{fig:dist_pr}
\end{figure}

\begin{figure*}[h]
    \centering
    \includegraphics[width = 1\linewidth]{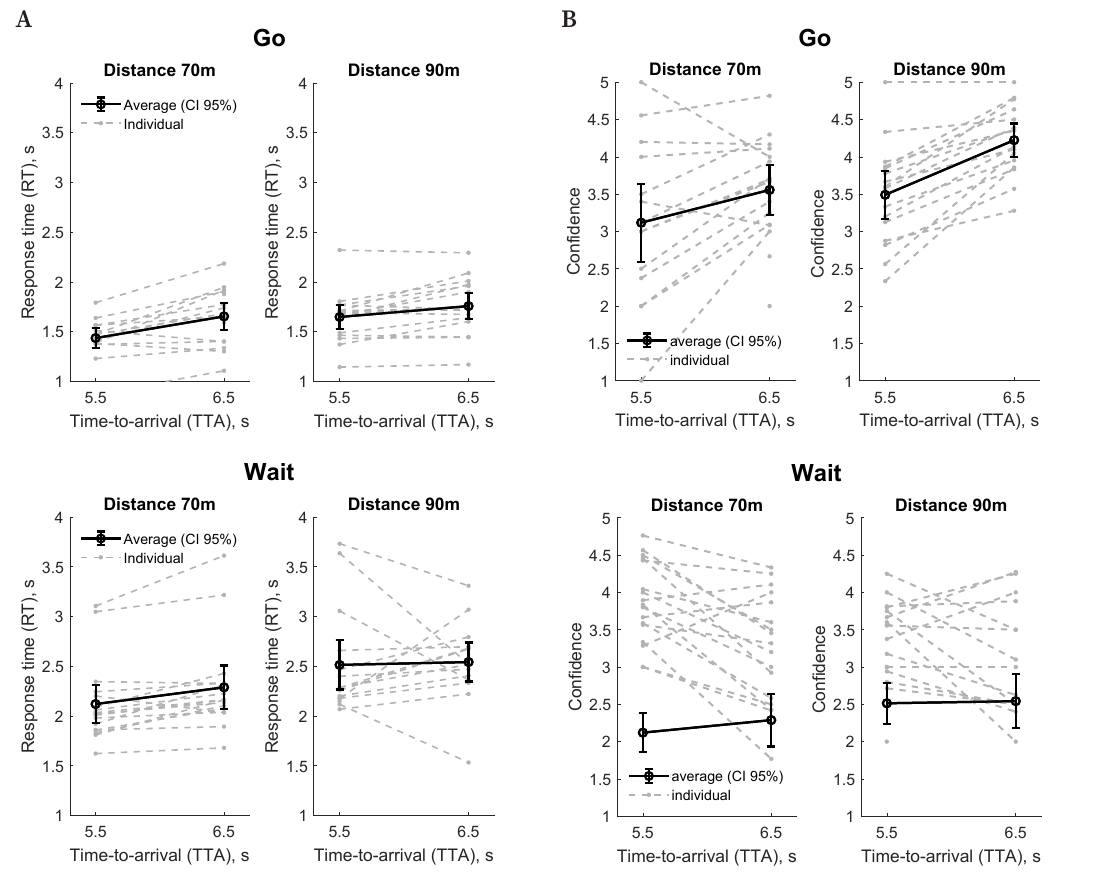}
    \caption{Group-averaged and individual mean A) response times, and B) confidence judgments for the two decision outcomes (``go''/ ”wait”) depending on the time-to-arrival and distance gap to the oncoming vehicle. Error bars represent 95$\%$ CI over $n = 17$ individual participants’ mean values.}
    \label{fig:dist_conf_rt}
\end{figure*}

We found that the drivers’ decision behavior (Figure~\ref{fig:dist_pr}, Table~\ref{tab:table_pr}) and response times (Figure~\ref{fig:dist_conf_rt}A, Table~\ref{tab:lme_rt}) were influenced by the time-to-arrival and distance gap of the oncoming vehicle. 

As expected (H1.1), we found that probability of making a ``go'' decision increased with time-to-arrival (\textit{t} = 9.2, \textit{p} = 1.7e-19) and distance gap (\textit{t} = 19.1, \textit{p} = 4.5e-73). 

\begin{table}[h!]
\centering
\begin{tabular}{@{}lllll@{}}
\toprule
            & Estimate & Std.   Error & t-score & pValue    \\ \midrule
(Intercept) & -2.175   & 0.1562       & -13.93  & 1.336e-41 \\
TTA         & 0.1860   & 0.02031      & 9.157   & 1.659e-19 \\
Distance    & 0.01940  & 0.001016     & 19.09   & 4.508e-73 \\ \bottomrule
\end{tabular}
\caption{Coefficients of the logistic regression describing the relation between the decision outcome and the TTA and distance gap (Eq.~\ref{eq:lme_pr})
. ``Wait'' decisions were coded as 0, ``go'' decisions as 1.}
\label{tab:table_pr}
\end{table}

We did not find evidence for a difference in mean response times between ``go'' and ``wait'' decisions ($b=-0.2, p=0.55$). Response times in both ``go'' and ``wait'' decisions increased with TTA and distance gap (Figure \ref{fig:dist_conf_rt}A, Table~\ref{tab:lme_rt}). For ``go'' decisions, response times were positively affected by the time-to-arrival ($b=0.12, t=4.0, p=6.5e-05$) and the distance gap ($b=0.007, t=4.4, p=9.9e-06$). We found no evidence that the effect of time-to-arrival was different in ``wait" decisions ($b=0.013, t=0.29, p=0.77$), but there was an additional positive effect of the distance gap on ``wait'' response times ($b=0.011, t=4.6, p=5.5e-06$) compared to ``go'' response times. (The detailed analyses of the mixed-effects models can be found in online supplementary information, Appendices A and B.)

We found that ``go'' response times we measured using button presses are moderately correlated with the response time suggested by the throttle input ($r=0.28, p=1.6e-15$), the measure that was previously suggested as a proxy measure of ``go'' response times~\cite{zgonnikov_should_2022,zgonnikov_subtle_2023}. However, in our data the latter often indicated deflated response times due to participants starting to press the throttle even before the could see the oncoming vehicle (see online supplementary information, Appendix C).

In short, we found that for both decision outcomes, response times were positively affected by the time-to-arrival (consistent with H1.2) as well as the distance gap (contrary to H1.2). In addition, the distance gap influenced the response time more for ``wait'' decisions than for ``go'' decisions.

\begin{table}[h!]
\centering
\begin{tabular}{@{}lllll@{}}
\toprule
                        & Estimate & Std. Error & t-score & pValue    \\ \midrule
Intercept               & 0.3325   & 0.2522     & 1.318  & 0.1876 \\
TTA                     & 0.1225   & 0.03059    & 4.005   & 6.502e-05 \\
Distance                & 0.007318  & 0.001651   & 4.434   & 9.934e-06 \\  \midrule
Wait decision           & -0.2091    & 0.3488     & -0.5996   & 0.5489  \\
Wait decision: TTA      & 0.01251   & 0.04327     & 0.2892  & 0.7725 \\
Wait decision: Distance & 0.01073  & 0.002353   & 4.560  & 5.525e-06 \\
\end{tabular}
\caption{Results of the linear mixed-effects regression analysis describing the effect of the TTA and distance gap on the response time (Eq.~\ref{eq:lme_RT}). ``Go'' was the reference category for the decision variable; the coefficients for the wait decisions represent additional effects with respect to the effects on ``go'' decisions.}
\label{tab:lme_rt}
\end{table}

\subsection{Confidence}

\subsubsection{Effect of time-to-arrival and distance gap on confidence}

Our results indicate that confidence reported in ``wait'' decisions was significantly higher compared to ``go'' decisions ($b=10.3, t=15, p=1.6e-48$); furthermore, in both decisions confidence depended on the time-to-arrival as well as the distance gap (Table~\ref{tab:conf_rt}, Figure~\ref{fig:dist_conf_rt}B).

For ``go'' decisions, the results displayed a positive relation between confidence on the one hand and the time-to-arrival ($b=0.74, t = 12, p = 1.8e-32$) and distance gap ($b= 0.034, t = 10, p = 1.85e-23$) on the other hand. 

For ``wait'' decisions, we observed, relative to ``go'' decisions, a negative relation between confidence and respectively the time-to-arrival ($b=-1.1, t=-13, p = 5.2e-37$) and the distance gap ($b=-0.04, t = -9.3, p = 7.1e-20$). Post-hoc comparisons revealed that this relative effect resulted in negative net effect of the time-to-arrival ($b = -0.38, F = 40, p = 3.9e-10$) and distance gap ($b = -0.011, F = 9.5, p = 0.0021$) on confidence in ``wait'' decisions. These findings indicate that the relation between confidence on the one hand and the time-to-arrival and the distance gap of the oncoming vehicle on the other is positive for ``go'' decisions and negative for ``wait'' decisions (consistent with H2.1).

\begin{table}[]
\centering
\begin{tabular}{@{}lllll@{}}
\toprule
                        & Estimate & Std. Error & t-score & pValue    \\ \midrule
Intercept               & -2.277   & 0.4975     & -4.576  & 5.126e-06 \\
TTA                     & 0.7413   & 0.06104    & 12.14   & 1.830e-32 \\
Distance                & 0.03361  & 0.003304   & 10.17   & 1.465e-23 \\ 
RT                      & -0.7411  & 0.08663    & -8.554  & 2.841e-17 \\ \midrule
Wait decision           & 10.26    & 0.6763     & 15.17   & 1.599e-48 \\
Wait decision: TTA      & -1.123   & 0.0886     & -13.06  & 5.231e-37 \\
Wait decision: Distance & -0.04432 & 0.004790   & -9.253  & 7.133e-20 \\
Wat decision: RT        & 0.1349   & 0.1055     & 1.279   & 0.2013   
\end{tabular}
\caption{Results of linear mixed-effects regression analysis of the effect of the response time, distance gap, and TTA on confidence judgments for different decision outcomes (Eq.~\ref{eq:lme_conf_rt}). The reference class in the regressions are the ``go'' decisions; the fixed effects coefficients for ``wait'' decisions are relative to the ``go'' decision coefficients.}
\label{tab:conf_rt}
\end{table}

\subsubsection{Relationship between response time and confidence}
We observed negative correlation between the response time and confidence ($r=-0.27, p=1.3e-27$) across all decisions. Mixed-effects regression (Eq.~\ref{eq:lme_conf_rt}) confirmed this negative relationship for ``go'' decisions ($b = -0.74, t = -8.6, p =2.84e-17$, see also Table~\ref{tab:conf_rt}). Furthermore, we observed no significant differences in the relationship of the response time and confidence between ``go'' and in ``wait'' decisions ($b=0.13, t=1.3, p=0.2$). Altogether, these findings substantiate the hypothesis that response time and confidence are negatively correlated (H3).

\subsubsection{Relationship between confidence and velocity profile}
\begin{figure*}
    \centering
    \includegraphics[clip, width = 1\linewidth]{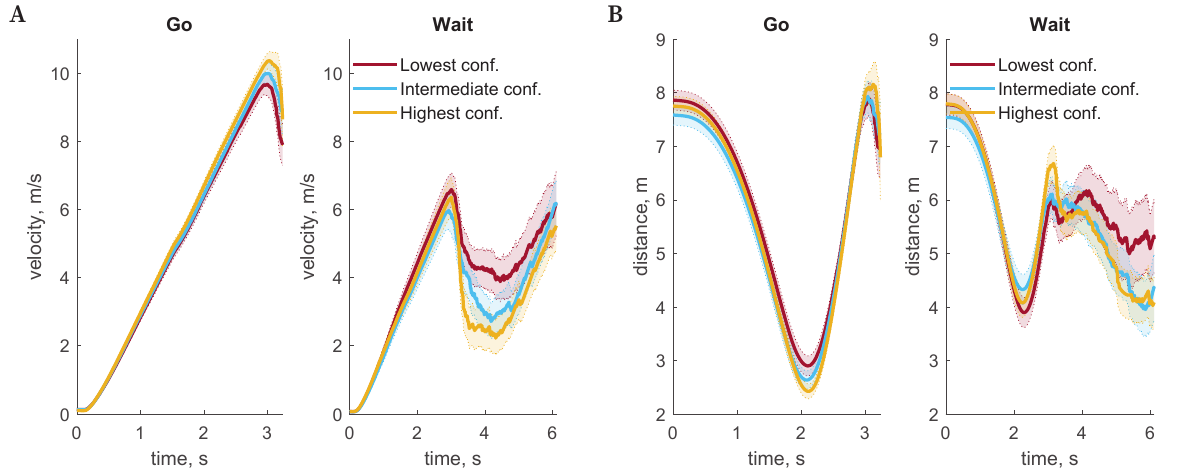}
    \caption{Group-averaged means values over time of participant's vehicle A) velocity, and B) distance to the centre of the intersection. For visualization purposes, trials were grouped in three equal-sized bins according to the reported confidence level (within decision outcomes); each line represents 1/3 of all ``go''/``wait'' trials. Shaded areas denote 95\% confidence intervals. Statistical analyses of action dynamics measures (Appendix C) suggested that the observed relationship between velocity profile and confidence is due to individual differences while the relationship between the distance to the center of the intersection and confidence persisted when controlling for individual differences.}
    \label{fig:vel_dist}
\end{figure*}

Visual inspection of group-averaged velocity profiles in ``go'' trials suggested
that in higher-confidence trials, participants seemed to exhibit higher maximum velocities (Figure~\ref{fig:vel_dist}A). However, statistical analyses controlling for individual differences in baseline metric values demonstrated that there was no evidence for a relationship between maximum velocity and confidence (``go'' trials: $b=0.1, t=1.2, p=0.22$; relative slope in ``wait'' trials: $b=-0.13, t=-1.2, p=0.22$). Similarly, there was no evidence for a relationship between confidence and average deviation from the individual mean, as well as RMSD (see online supplementary information, Appendix E). At the same time, there was a weak positive relation between the deviation from the group mean of the velocity profile and confidence in ``go'' trials ($b= 0.11, t= 2.2, p= 0.03$), the effect that was not significantly different in ``wait'' trials ($b=-0.09, t= -1.4, p= 0.16$).

Taken together, these analyses suggest no evidence for an underlying relationship between confidence and velocity profile within participants (contrary to H4.1a, H4.2a, and H4.3a\footnote{Testing whether relationship between confidence and velocity profiles was weaker in ``wait'' decisions compared to ``go'' decisions (H4.4) was not meaningful due to the lack of evidence of such relationships in ``go'' decisions in the first place.}); yet participants who were more likely to report higher confidence levels also exhibited slightly higher velocities in ``go'' decisions. 

The latter finding, however, should be interpreted with caution, given the relatively small effect size.

\subsubsection{Relationship between confidence and distance to the centre of the intersection} 
Dynamics of mean distance over time (Figure~\ref{fig:vel_dist}B) indicates a possible negative relation between confidence and the minimum distance to the intersection centre. Similar to the velocity profiles, this could be because higher confidence judgments are associated with more pronounced corner-cutting behavior within or across participants. 
Mixed-effects regression analyses confirmed a relationship between reported confidence and the distance to the centre of the intersection (see online supplementary information, Appendix E). Specifically, in ``go'' decisions, confidence was associated with increased corner-cutting behavior, i.e. had a negative relation with the minimum distance to intersection centre ($b=-0.17, t = -3.9, p = 8.5e-5$) as well as with the deviation from the group mean ($b=-0.11, t = -2.4, p = 0.015$) and RMSD from the individual mean ($b=-0.07, t=-2.7, p=0.006$), although there was no evidence for its relationship with the signed average deviation from the individual mean ($b=-0.04, t=-1.1, p=0.28$). These effects (except for the latter one) were significantly different in ``wait'' decisions. Post-hoc comparisons highlighted the lack of evidence for a relationship between confidence reported in ``wait'' decisions and minimum distance from the centre of the intersection ($b= -0.73, F = 0.41, p = 0.52$), average deviation from its group mean ($b=-0.64, F = 2.5, p = 0.12$), and RMSD from the individual mean ($b=0.06, F =  1.5, p = 0.23$).

Overall, our analyses of the vehicle trajectories following a decision confirmed that for ``go'' decisions, the minimum distance but not average distance to the intersection centre related negatively to confidence within participants (consistent with H4.1b but contrary to H4.2b). This finding hints towards more pronounced corner-cutting behavior after high-confidence ``go'' decisions. In addition, the results highlight increased variability in turning trajectories after lower-confidence decisions (consistent with H4.3b).

Finally, our analyses suggest that action dynamics after ``wait'' decisions are not related to reported confidence levels, which is in line with H4.4 that decision confidence mainly affects ``go'' decisions.

\section{Cognitive process modelling of confidence judgments}

\begin{figure*}
    \centering
    \includegraphics[width = 1\linewidth]{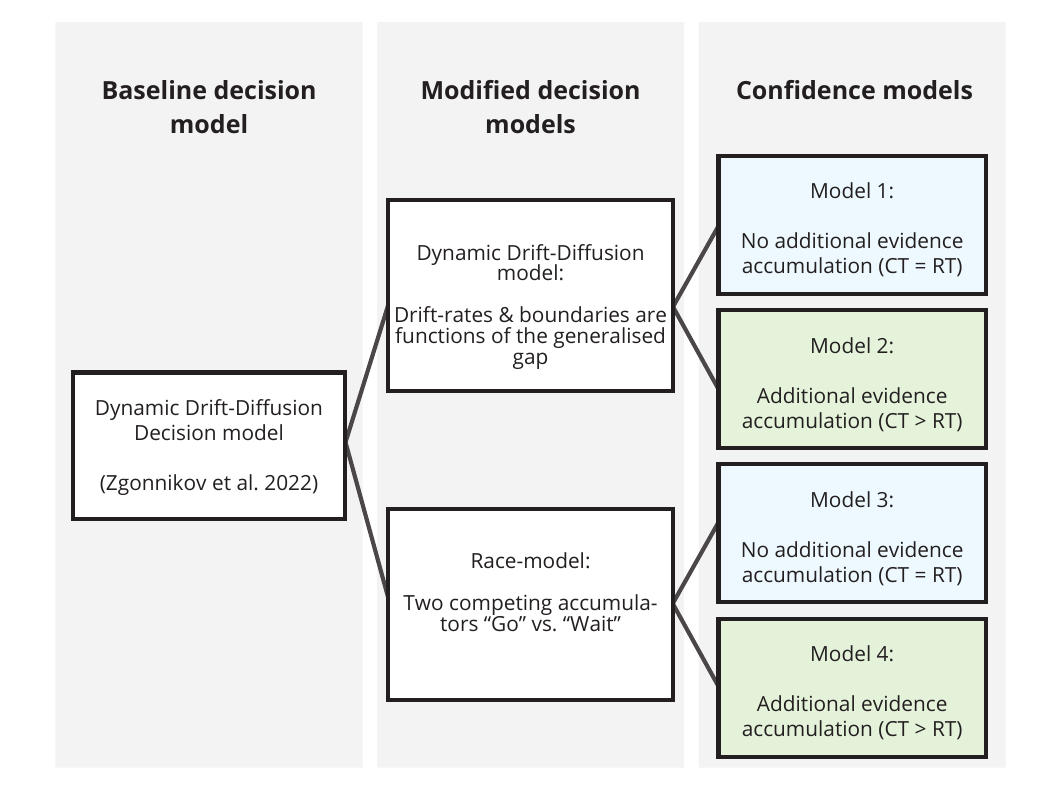}
    \caption{Four tested confidence models derived from combining two candidate decision-making mechanisms (drift-diffusion model or race model) and two different time points of determining the confidence judgment (at the moment of making the decision or after accumulating additional evidence). CT: confidence judgment time, RT: response time.}
    \label{fig:model_overview}
\end{figure*}

To elucidate the cognitive processes underlying the response times and confidence judgments observed in our experiment, we tested four potential confidence models (Figure~\ref{fig:model_overview}). All four models are grounded in the evidence accumulation framework, but differ in their assumptions about the underlying decision-making mechanism (drift-diffusion model~\cite{ratcliff_diffusion_2008} vs. race model~\cite{bogacz_physics_2006}) and the confidence judgment process (post-decision accumulation vs. judgment based on evidence accumulated prior to decision).

\subsection{Decision-making models}
Both candidate decision-making models were based on a previously suggested model of left-turn gap acceptance~\cite{zgonnikov_should_2022}. This model is a variant of the drift-diffusion model~\cite{ratcliff_diffusion_2008} which assumes that the decision-making process can be conceptualized as noisy evidence accumulation
\begin{equation} \label{eq: dx_basis}
    dx(t) = \alpha(g(t) -\theta_{\textrm{crit}})dt + dW,
\end{equation}
where $x(t)$ is the decision variable at time $t$, $W$ is the stochastic Wiener process, and $\alpha(g(t) -\theta_{\textrm{crit}})$ is the rate of evidence accumulation. The parameter $\alpha \geq 0$ indicated the influence of the perceptual information on the evidence accumulation process relative to the noise. The model assumed that the rate of evidence accumulation is proportional to the difference between the ``generalized gap'' $g(t)$ and the critical value of the generalised gap $\theta_{\textrm{crit}}\geq0$. The generalized gap $g(t)$ was hypothesized to be a linear combination of the time-to-arrival and the distance gap to the oncoming vehicle 
\begin{equation} \label{eq:generalised_gap}
    g(t) = TTA(t) + \beta d(t),
\end{equation}
with $\beta \geq 0$ characterizing the relative contribution of the distance gap relative to the time-to-arrival.

The moment in time when $x$ reaches one of the decision boundaries ($\pm b(t)$) is defined as the response time; the upper boundary corresponds to the decision to accept the gap (``go") and the lower boundary represents the decision to reject the gap (``wait"). The boundaries were assumed to be collapsing with the time-to-arrival of the oncoming vehicle:
\begin{equation} \label{eq: b_basis}
    b(t) = \frac{b_0}{(1+e^{-k(TTA(t)-\tau)})},
\end{equation}
where parameters $b_0>0$, $k\geq0$ and $\tau\geq0$ indicated respectively the boundary scale parameter, the sensitivity of the boundary to the time-to-arrival, and the characteristic time-to-arrival value at which the boundary is equal to $\pm\frac{1}{2}b_0$. Finally, the model accounted for possible perceptual and response delays via a normally-distributed non-decision time
\begin{equation} \label{eq:non_dec_t}
    t_{\textrm{ND}}=N(\mu_{\textrm{ND}},\sigma_{\textrm{ND}})
\end{equation}
 with mean $\mu_{\textrm{ND}}\geq0$ and standard deviation $\sigma_{\textrm{ND}}\geq0$.
 
In this paper, we propose two modified versions of the baseline model: the drift-diffusion model (Figure~\ref{fig:ev_example}A) and the race model which describes the decision-making process with two independent accumulators (Figure~\ref{fig:ev_example}C). Both proposed models describe the decision process with generalized gap-dependent collapsing boundaries.

\subsubsection{Drift-diffusion model}
The baseline model assumed that the decision boundaries collapse with the time-to-arrival, which was based on the observations of Zgonnikov et al.~\cite{zgonnikov_should_2022} that response time increased with time-to-arrival but not distance. In our study however, we found that response time also increased with the distance to the oncoming vehicle. In order for the model to be able to exhibit such behavior, we assumed that the decision boundary collapses not with the time-to-arrival, but with the generalized gap (which incorporates both time-to-arrival and distance gap): 
\begin{equation}\label{eq:b_new}
    b(t)=\pm \frac{b_{0}}{(1+e^{-k(g(t)-\theta_{\textrm{crit}})})},
\end{equation}
where $g(t)$ is described by Equation~\eqref{eq:generalised_gap}. 

Besides the decision boundary, all the other components of the baseline model were retained for the modified drift-diffusion model. This model has seven free parameters ($\alpha$, $\beta$, $\theta_{\textrm{crit}}$, $b_0$, $k$, $\mu_{\textrm{ND}}$, $\sigma_{\textrm{ND}}$)%

\subsubsection{Race model}
\begin{figure*}[t!]
    \centering
    \includegraphics[width = 1\linewidth]{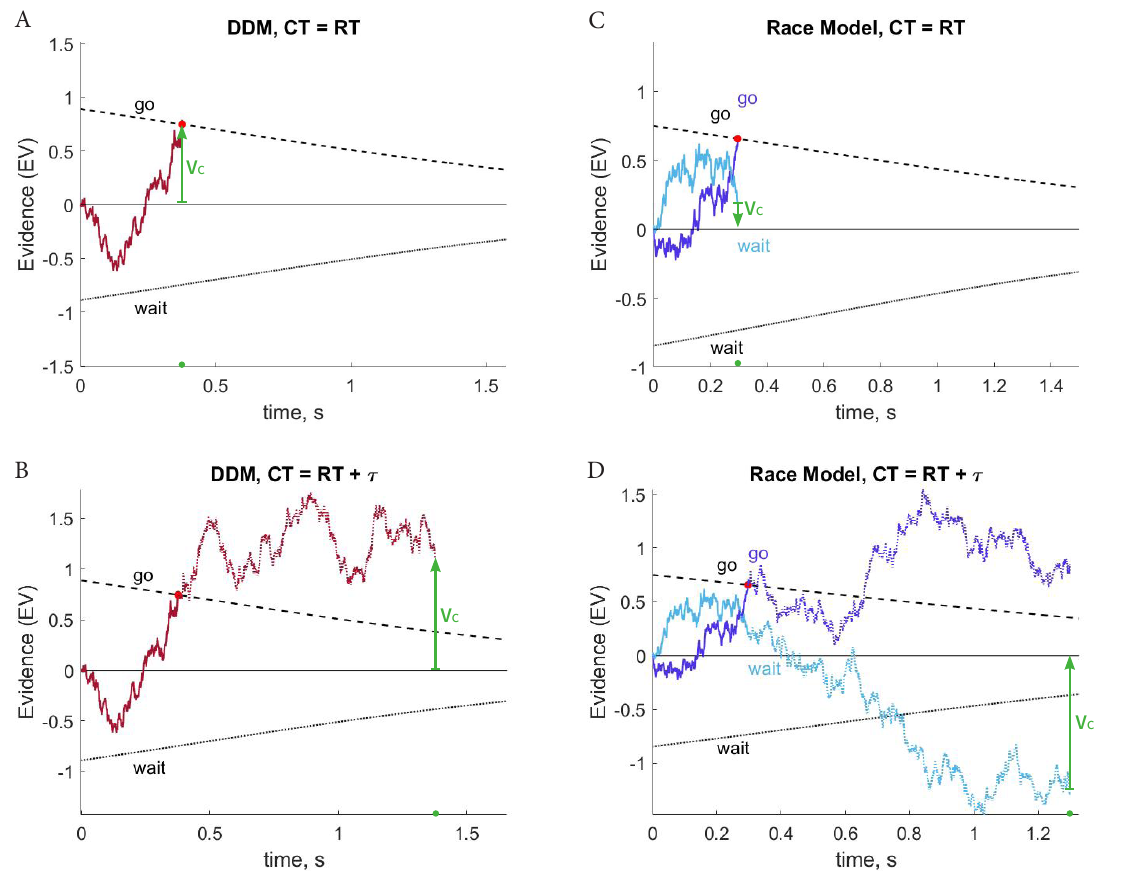}
    \caption{Illustration of the four tested models. In each panel, the red dot indicates the decision moment (RT) and the green dot indicates the moment in time the confidence judgment is made (CT). The green arrow represents the value of the confidence variable $V_c$ that is used for the computation of the confidence judgment (Eq.~\ref{eq:Conf}). 
    }
    \label{fig:ev_example}
\end{figure*}

The race model (Figure~\ref{fig:ev_example}C) included two independent competing decision variables, $x_{\textrm{go}}$ and $x_{\textrm{wait}}$, which are responsible for the decision outcome~\cite{bogacz_physics_2006, gold_neural_2007, kiani_choice_2014}. The dynamics of both decision variables are based on the same perceptual evidence (the generalized gap $g(t)$, see Equation~\eqref{eq:generalised_gap})

\begin{equation} \label{eq:dx_race_go}
    dx_{\textrm{go}}(t) = \alpha_{\textrm{go}}(g(t) - \theta_{\textrm{crit}})dt + dW_{\textrm{go}},
\end{equation}
\begin{equation} \label{eq:dx_race_wait}
    dx_{\textrm{wait}}(t) = -\alpha_{\textrm{wait}}(g(t)-\theta_{\textrm{crit}})dt + dW_{\textrm{wait}}.
\end{equation}
Here, we assumed that the drift rate $\alpha$ could differ between the two accumulators while the parameters $\beta$ and $\theta_{\textrm{crit}}$ defining the generalized gap $g(t)$ and its characteristic value, respectively, remain the same for the two accumulators. The reasoning underlying this assumption is that ``go'' decisions involve more risk and therefore could require more efficient accumulation of evidence, while the drivers’ perception of the generalised gap is likely to be independent from the decision. 

In the decision boundary, the scale parameter $b_0$ as well as the collapse rate $k$ were also assumed to be different for ``go'' and ``wait'' accumulators 

\begin{equation} \label{eq:b_race_go}
    b_{\textrm{go}}(t)= \frac{b_{0}^{\text{go}}}{(1+e^{-k^{\text{go}}(g(t)-\theta_{\textrm{crit}})})},
\end{equation}
\begin{equation} \label{eq:b_race_wait}
    b_{\textrm{wait}}(t)= \frac{b_{0}^{\text{wait}}}{(1+e^{-k^{\text{wait}}(g(t)-\theta_{\textrm{crit}})})}.
\end{equation}

The non-decision time was assumed to be the same as in the baseline model (Equation~\eqref{eq:non_dec_t}), and was identical for the two accumulators. Overall, our race model has ten free parameters ($\alpha_{\textrm{go}}$, $\alpha_{\textrm{wait}}$, $\beta$, $\theta_{\textrm{crit}}$, $b_0^{\text{go}}$, $b_0^{\text{wait}}$, $k^{\text{go}}$, $k^{\text{wait}}$, $\mu_{\textrm{ND}}$, $\sigma_{\textrm{ND}}$).

\subsection{Confidence models}
The different confidence models were based on the two alternative accounts of how evidence accumulation is used in confidence judgments. The first account states that the amount of evidence accumulated at the time the decision is made determines the confidence judgment~\cite{kiani_choice_2014,meyniel_schluneger_2015,Boundy_Singer2022}. The second account argues that confidence judgments instead invoke accumulating additional evidence post-decision~\cite{murphy_neural_2015, yeung_metacognition_2012, pleskac_two-stage_2010, kobe2020}. We explored possible combinations of these two accounts of confidence judgment time with the two candidate decision models, as defined in the previous section. This resulted in four potential confidence models (Figure \ref{fig:model_overview}). 

The four tested confidence models shared the basic premise that the evidence accumulation process responsible for the decision-making process also plays role in confidence judgments. In all models, the confidence judgment $c$ was hypothesized to be a linear function of the confidence variable $V_c(t)$ evaluated at the confidence judgment time (CT)
\begin{equation}\label{eq:Conf}
    c = c_{0}^{\textrm{dec}} + c_{1}^{\textrm{dec}}V_c(\textrm{CT}).
\end{equation}
Here, the intercept $c_{0}^{\textrm{dec}}$ and the sensitivity parameter $c_{1}^{\textrm{dec}}>0$ map the decision variable to the confidence rating scale (in the case of our experiment, 1 to 5). Both of these parameters depend on the decision, reflecting our finding that for ``go'' decisions, confidence judgments are differently biased and influenced in a different manner by perceptual information than for ``wait'' decisions.

The four models differed in the manner in which they queried the evidence accumulation process to determine the confidence variable $V_c(t)$ and the moment at which the confidence variable is queried (CT).

\subsubsection{Confidence variable $V_c(t)$}
Models 1 and 2 were based on the drift-diffusion model and models 3 and 4 were based on the race model; the underlying decision model determined the manner in which the confidence variable $V_c(t)$ was calculated. 

The drift-diffusion model-based confidence models (model 1 and 2) had one decision variable denoting the relative evidence (Figure~\ref{fig:ev_example}B). These models hypothesized that the confidence variable is equal to the absolute value of this relative evidence ($V_c(t)=|x(t)|$), or alternatively
\begin{equation} \label{eq:Vc_DDM}
    V_c(t)=  \left\{ \begin{matrix}x(t) & \textrm{if decision = ``Go''} \\ -x(t) & \textrm{if decision = ``Wait''} \end{matrix}\right.	
\end{equation} 

The race-model-based confidence models (model 3 and 4) used two accumulators, and posited that the confidence judgment depends on the value of the evidence accumulated towards the non-chosen (alternative) option (Figure~\ref{fig:ev_example}D). For these models, the confidence variable was defined as the negative value of the alternative evidence accumulator 
\begin{equation} \label{eq:Vc_race}
    V_c(t)=  \left\{ \begin{matrix} -x_{\textrm{wait}}(t) & \textrm{if decision = ``Go''} \\ -x_{\textrm{go}}(t) & \textrm{if decision = ``Wait''} \end{matrix}\right.
\end{equation}

\subsubsection{Confidence judgment time $\textrm{CT}$}
Models 1 and 3 assumed that the confidence judgment is made based on the values of the decision variable(s) at the time the decision is made, whereas models 2 and 4 allowed for post-decision evidence accumulation.

The models without post-decision evidence accumulation (models 1 and 3) assumed that the confidence response time is equal to the response time 
\begin{equation}\label{eq:CT_no}
    \textrm{CT} = \textrm{RT}
\end{equation} 
This implies that the same value(s) of the evidence accumulator(s) are used for both the decision-making process and for the confidence judgment. For the drift-diffusion model, this implies that confidence relates to the value of the decision boundary at the response time. As in the underlying drift-diffusion model the decision boundary collapses with time, the confidence in model 1 depends on the response time~\cite{kiani_choice_2014}. For the race model, this is not the case as the losing accumulator defines the confidence variable $V_c(t)$. Hence, in model 3, confidence depends on the evidence accumulated by the losing accumulator at the response time. 

The models with post-decision evidence accumulation (models 2 and 4), the evidence accumulation process continues for  $\tau>0$ seconds after the decision is made 
\begin{equation} \label{eq:CT_ad}
    \textrm{CT} = \textrm{RT} + \tau.
\end{equation}
As a result, the value(s) of the evidence accumulator(s) used for the decision making differ(s) from the value(s) used for confidence judgments. This means that in models 2 and 4 the confidence judgments depend on the evidence accumulated by the single accumulator (model 2) or the losing accumulator (model 4) at $\tau$ seconds after the response time.

\subsection{Model fitting}
We used a two-stage approach for model fitting (see online supplementary information, Appendix F for details). First, we fitted the decision model parameters ($\alpha$, $\beta$, $b_0$, k, $\mu_{\text{ND}}$, $\sigma_{\text{ND}}$, and $\theta_{crit}$) to the decision outcome and response time data, following the weighted least squares approach~\cite{Ratcliff_tuerlinckx_2002, zgonnikov_should_2022}. Second, using the baseline decision model parameters fitted in the first stage, we fitted the confidence model parameters (sensitivity, $c_1$ and bias, $c_0$) using the root mean square error of model confidence outputs. For both stages of  model fitting, ``fmincon'' function of MATLAB R2020a was used to find the best-fitting parameters. 

For the confidence models that involved post-decision evidence accumulation, we did not include the parameter $\tau$ in the second model fitting stage. The reason for this was that changes in this parameter did not substantially affect model behavior (see online supplementary information, Appendix F). In both models with post-decision accumulation we hence used $\tau=1s$ as inter-judgment time. To put things in perspective, the average duration of making the turn (time between the first moment of seeing the oncoming vehicle and giving the confidence judgment) was 3.06 seconds (\textit{SD} = 0.498) for “go” decisions and 6.26 seconds (\textit{SD} = 1.092) for “wait” decisions. 

\subsection{Results}
\begin{figure*}[t!]
    \centering
    \includegraphics[width = 1\linewidth]{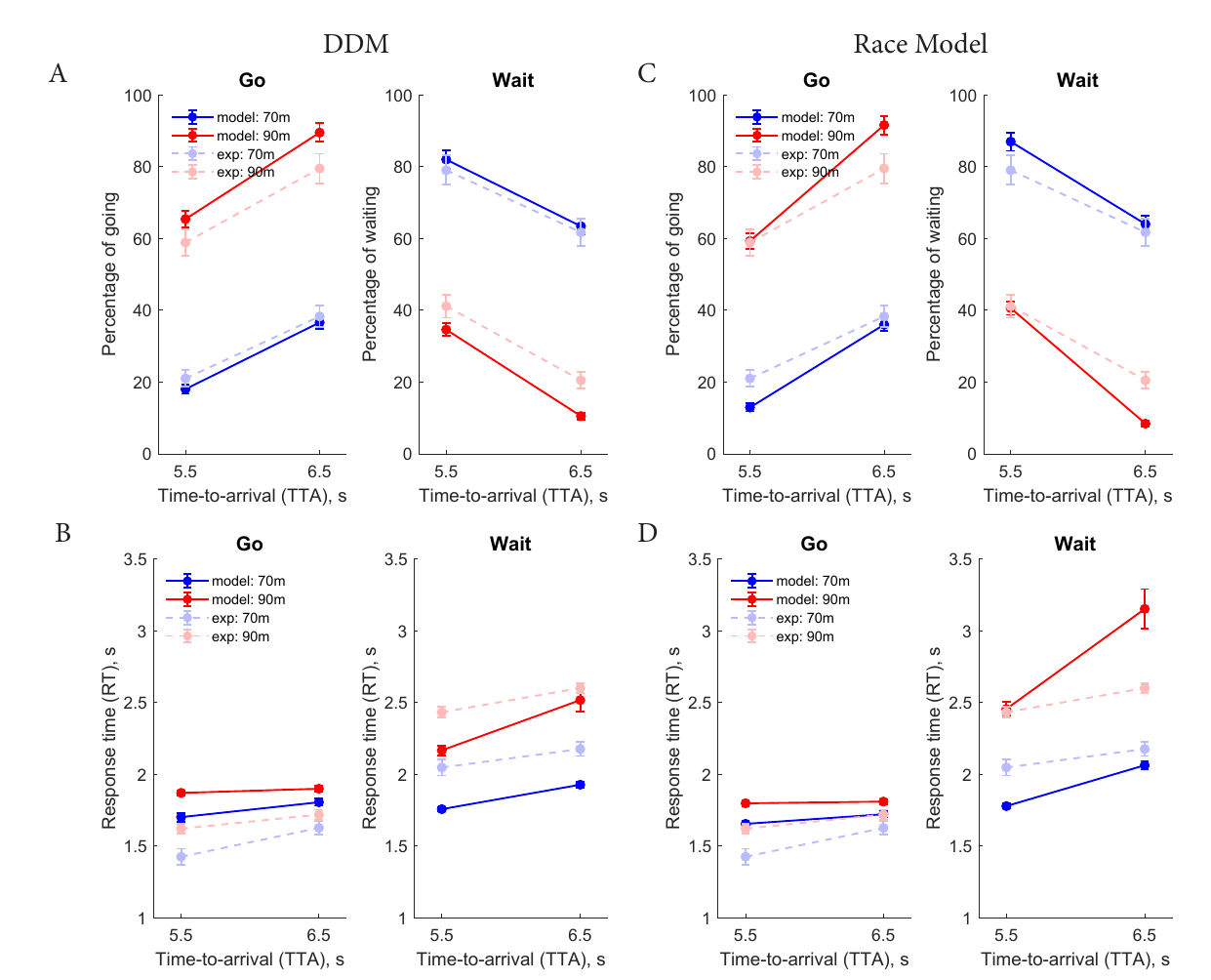}
    \caption{Decision outcomes (A, C) and response times (B, D) produced by the drift-diffusion model (A, B) and the race model (C, D). Error bars: 95\% CI. 
    }
    \label{fig:decision_model_comparison}
\end{figure*}

We found that both the DDM and the race model had similar performance in fitting the experimentally observed decisions and response times (Figure~\ref{fig:decision_model_comparison}, Table~\ref{tab:model_pr}). The race model performed slightly better in terms of WLS, although this should be taken with caution given a larger number of free parameters in this model (10 in race model vs 7 in DDM).

\begin{table*}[!t]
\centering
\begin{tabular}{@{}l|l|llllllll@{}}
\toprule
 & WLS & $\alpha$ & \multicolumn{2}{l}{$\beta$} & $b_0$ & $k$ & $\mu_{ND}$ & $\sigma_{ND}$ & $\theta_{crit}$ \\ \midrule
DDM & 1.53 & 1.12 & \multicolumn{2}{l}{0.109} & 1.41 & 0.396 & 1.51 & 0.140 & 14.0 \\
Race & 1.45 & \begin{tabular}[c]{@{}l@{}}Go:   1.20  \\ Wait:   1.14\end{tabular} & \multicolumn{2}{l}{0.100} & \begin{tabular}[c]{@{}l@{}}Go:  1.22 \\
Wait: 1.35
\end{tabular} & \begin{tabular}[c]{@{}l@{}}Go: 0.398 \\ Wait: 0.438 \end{tabular}& 1.49 & 0.120 & 13.3 \\ \bottomrule
\end{tabular}
\caption{Fitted parameters of the drift-diffusion and race decision models; WLS: mean weighted least squares error~\cite{Ratcliff_tuerlinckx_2002}.}
\label{tab:model_pr}
\end{table*}

The fitted parameters of the race model show that the drift rate $\alpha$ was slightly higher for the ``go" accumulator, as compared to the ``wait" accumulator. Conversely, the baseline boundary $b_0$ was higher for the ``wait" accumulator compared to ``go", indicating that a larger amount of evidence needed to be accumulated to make a ``wait" decision. Finally, the decision boundary was collapsing slightly faster (as indicated by a larger value of parameter $k$) for the ``wait" accumulator, as compared to the ``go" accumulator. 

\begin{figure*}[h!]
    \centering
    \includegraphics[clip, width = 1\textwidth]{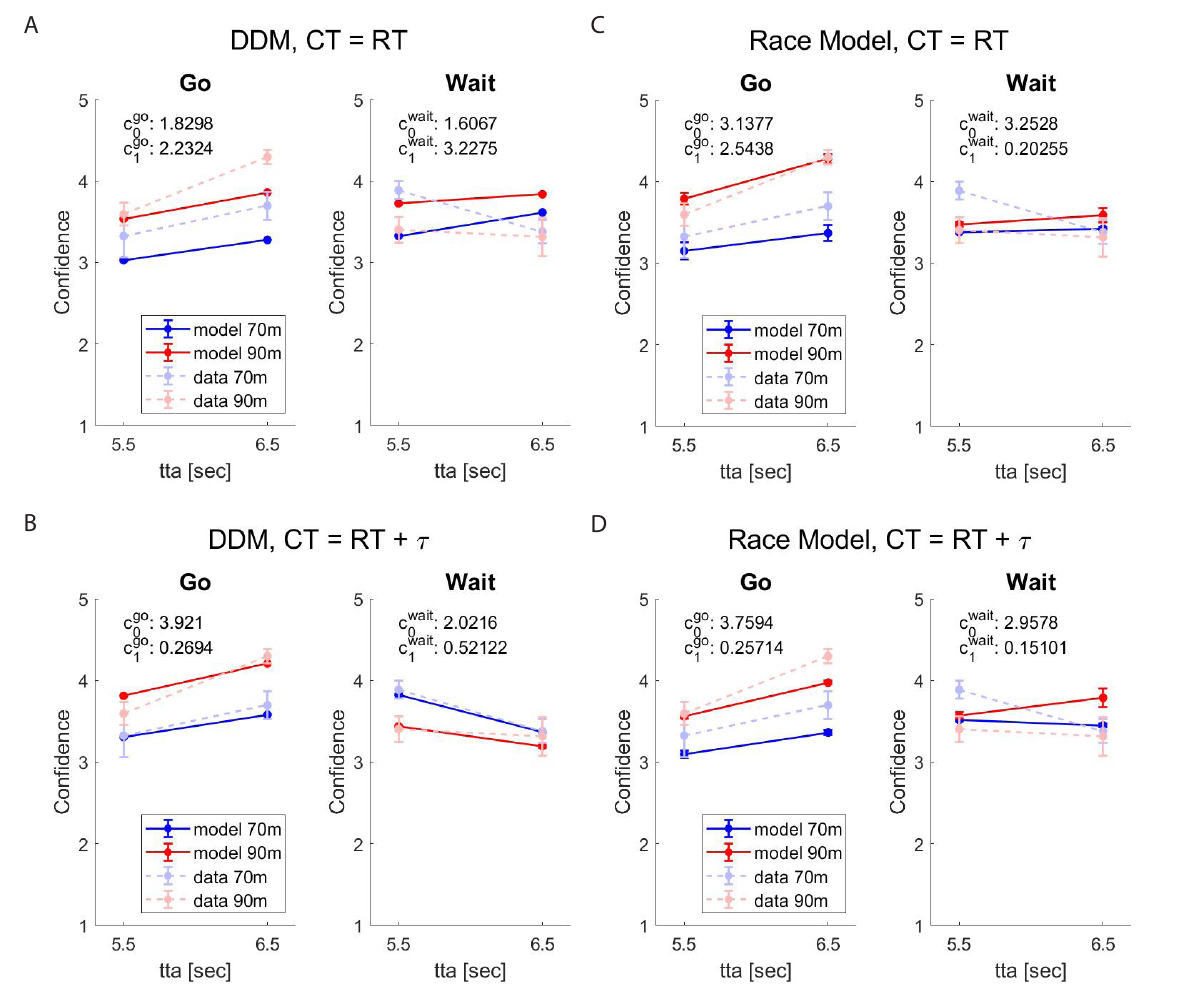}
    \caption{Confidence judgments produced by the four candidate models.}
    
    \label{fig:overview_conf}
\end{figure*}

Fitting of two candidate confidence models for each baseline decision models demonstrated that experimentally observed confidence ratings were more consistent with ratings produced after additional evidence accumulation rather than ratings measured at the time of the decision (Figure~\ref{fig:overview_conf}, Table~\ref{tab:conf_perform}). The additional evidence accumulation in particular improved the ability of the models to capture confidence judgments in “wait” decisions, for both the race model as well as the DDM. 

This finding substantiates the account of confidence judgments being based on post-decision evidence accumulation~\cite{yeung_metacognition_2012,fleming_self-evaluation_2017, murphy_neural_2015}.

\begin{table}[h!]
\begin{tabular}{@{}l|ll@{}} \toprule
                                   & DDM             & Race model     \\ \midrule
CT = RT          & Model 1: 0.403  & Model 3: 0.258 \\
CT = RT + $\tau$ & Model 2: 0.107 & Model 4: 0.274  \\ \bottomrule
\end{tabular}
\caption{Performance of the four tested confidence models: \textit{RMSE} relative to the confidence ratings measured in the experiment.}
\label{tab:conf_perform}
\end{table}

Furthermore, we observed that the confidence judgment process was best described using the dynamic drift-diffusion decision model (model 2). 
This implies that confidence judgments are better described by one evidence accumulator accounting for both decision outcomes, instead of two independent evidence accumulators for both decisions as described by the race model. 

Altogether, our modelling results indicate that observed confidence judgments were most consistent with the drift-diffusion model that allows for post-decision evidence accumulation. 
\section{Discussion}\label{sec:discussion}
 Previous studies of simple perceptual and preferential tasks provided evidence that the cognitive processes responsible for decisions and confidence judgments are closely related to each other ~\cite{kiani_representation_2009,kepecs_computational_2012,pouget_confidence_2016,zylberberg_construction_2012}. However, computational underpinnings of confidence judgments in dynamic real-life tasks have remained unclear. Here we aimed to address this gap, focusing on left-turn gap acceptance decisions by human drivers.

\subsection{Relation to previous work}
\subsubsection{Response times in gap acceptance decisions}
Response time has been long established as a key metric in decision-making research, providing a basic measure of the time course of the decision-making process~\cite{luce_response_1986}. We found a positive relation between response time on the one hand and the distance gap and time-to-arrival on the other hand. These findings are consistent with the previous literature that reported such positive relationships in similar decisions during left turns across path~\cite{zgonnikov_should_2022,zgonnikov_subtle_2023}, overtaking~\cite{sevenster_response_2023,mohammad_cognitive_2023}, merging~\cite{zgonnikov_now_2023}, and lane changing~\cite{yan_investigating_2023}.

\subsubsection{Empirical studies of confidence}
Earlier studies of confidence in simple perceptual decisions have universally reported that a higher quality as well as a higher quantity of evidence towards a chosen option result in higher confidence judgments~\cite{kiani_choice_2014,yeung_metacognition_2012,murphy_neural_2015}. In the context of our task, this means that for example, the combination of a large time-to-arrival and a large distance gap should result in higher-confidence ``go'' decisions than e.g. a combination of a large time-to-arrival and a small distance gap. 
The results confirmed that in conditions with larger time-to-arrival and distance to the oncoming vehicle, participants reported higher confidence in “go” decisions and lower confidence in ``wait'' decisions. This corresponds to the commonly observed ``folded X-pattern" in perceptual decisions~\cite{kepecs_computational_2012, drugowitsch_becoming_2016, sanders_signatures_2016}. Furthermore, the negative relation we found between confidence and response time is also in accordance with earlier confidence research, which argued that less (qualitative) evidence towards a decision results in a longer duration of evidence accumulation before making a decision, which results in a reduced amount of confidence~\cite{kiani_choice_2014,pleskac_two-stage_2010}.
Our findings thus connect the domain of driver decision making to previous findings in the basic empirical literature on confidence. 

Among the existing applied work on driver decision making, the studies most related to this paper are the ones that studied drivers' \textit{certainty} (or, alternatively, uncertainty) in gap acceptance decisions in lane changes~\cite{yan_classifying_2015, yan_developing_2016, yan_investigating_2023}, narrow passages~\cite{miller_time_2022}, and overtaking~\cite{leitner_overtake_2023}.  
While closely related, the concepts of confidence and certainty are however fundamentally distinct, with known dissociations between them~\cite{peterson_confidence_1988, pouget_confidence_2016}. Our findings are generally consistent with the patterns of certainty judgments observed in previous studies of driver behavior~\cite{yan_classifying_2015, yan_developing_2016, yan_investigating_2023, miller_time_2022, leitner_overtake_2023}, although the manner in which (un)certainty is analyzed in the literature complicates a direct comparison. Specifically, in this line of research, (un)certainty has been analyzed as a function of kinematic variables (such as TTA) disregarding the decision outcome. This obscures the functional relationship between the evidence in a given option and the (un)certainty judgments in decisions in favor of that option: ``the key property of confidence is that it is choice dependent''~\cite{pouget_confidence_2016}. By addressing this issue, as well as testing specifically for dissociations previously reported for perceptual decisions~\cite{peterson_confidence_1988}, future work can clarify the relationship between certainty and confidence in driver decision making.

The main focus of this work is on subjective confidence reports, which continue to be a cornerstone of the fundamental literature on confidence. More recently, post-decision response trajectories have been suggested as a alternative measure of confidence~\cite{van_den_berg_common_2016, dotan_-line_2018}. These trajectories  could hypothetically provide a less intrusive and more dynamic alternative to traditional confidence measures. However, existing studies on the relationship between confidence and action dynamics have relied on highly specialized tasks that were specifically designed to probe this relationship. Our driving task offered us a unique opportunity to investigate how confidence relates to post-decision action dynamics (namely, vehicle trajectories) in a naturalistic setting. We found evidence for a relationship between confidence and some of the action dynamics measures (e.g., corner-cutting behavior) but not others (e.g., velocity). This suggests that potential ``leakage'' of confidence into post-decision behavior in naturalistic driver behavior is real but nuanced; follow-up studies measuring both subjective confidence reports and full driving trajectories in other driving tasks can further illuminate this relationship.

\subsubsection{Cognitive process modelling of confidence judgments}
We showed that the cognitive process responsible for confidence judgments can be modelled with the same evidence accumulator that is used to model the decision-making process when accounting for additional evidence accumulation. In order to translate the present evidence expressed by the evidence accumulator into a confidence judgment, we defined two additional confidence model parameters that describe the sensitivity towards the present evidence and the confidence bias. These additional model parameters can potentially be related to the two main factors that determine someone’s metacognitive ability: the metacognitive sensitivity representing the reliability or accuracy, and the metacognitive bias which can be described as the calibration~\cite{fleming_how_2014}. This can be used to explain the observed differences between individuals, something that may be of interest for future research. 

Our modelling results are consistent with previous research by Zylberberg et al~\cite{zylberberg_construction_2012} who found that post-decision confidence is more strongly affected by positive evidence in favour of the made decision than by the negative evidence in favour of the alternative decision in a simple perceptual task. More generally, our results contribute to the body of evidence that implicate post-decision evidence accumulation as a mechanism contributing to confidence judgments~\cite{pleskac_two-stage_2010, yeung_metacognition_2012, hellmann_simultaneous_2023}. We tested four candidate models, each of which was adapted from previous research; in that sense, our study did not yield any insights into previously unknown confidence mechanisms. Instead, our original contribution lies in evaluating the basic computational accounts of confidence in a close-to-real-life driving setting (something that hasn't been addressed in driver behavior literature so far). At the same time, more intricate models have recently been proposed which are yet to be tested in such settings~\cite{fleming_self-evaluation_2017, atiya_neural_2019, hellmann_simultaneous_2023}; we see this as yet another important direction for future work.

\subsection{Limitations}

In this study, we measured confidence with post-decision self-reports. The main advantage of using a self-report-based measurement method is that confidence judgments are directly measured~\cite{kepecs_computational_2012}. However, the disadvantage of this method, which has not been taken into account in this study, is that measurement errors and strategic biases can cause distortion in the measurement~\cite{kepecs_computational_2012}. For example, participants may not be willing to give an honest confidence judgment because of perceived expectations.
Furthermore, we only measured confidence after the performance of the turn. This makes it difficult to judge whether participants' reflection on their turn execution affected the confidence rating, despite the fact that we asked to report the confidence they initially had when making the the decision. 

Our modelling results suggest that participants were not able to do so, showing that confidence judgments were best explained when taking into account additional evidence accumulation. In other words, participants continued  evidence accumulation after the decision was made which affected their retrospective confidence judgment. Further research could make a comparison between self-reported confidence at the decision response time and self-reported confidence at the end of the turn to provide further insight on this issue.

The results of this study demonstrated that the confidence of a driver in left-turn gap accepting decisions is affected by the time-to-arrival and distance gap of the oncoming vehicle, besides individual differences. We made use of the aggregate data for the regression analysis and allowed for individual differences with random intercepts. However, we did not investigate or account for the causes of these individual differences in our cognitive modelling. Earlier research has shown that individual differences can be caused by differences in self-concept, the level of metacognitive skills, the level of skills and (the amount of) experience~\cite{handel_individual_2020,chua_building_2015,kruger_unskilled_noyear}.

For the reason of simplicity, we assumed in our post-decision evidence accumulation confidence models that the drift rate remains the same after the decision has been made. However, it can be questioned to what extent this assumption is true. For instance, one may argue that the contribution of new perceptual evidence reduces after the decision is made, especially in ``go'' decisions because the driver does not observe the relevant perceptual information anymore. Follow-up research could explore how the evidence accumulation after the decision has been made can best be described.

Lastly, the two-stage model fitting approach we adopted rested on the assumption that the decision-making and  confidence-judgment parts of the model are independent from each other, and therefore could be fitted to the data in a sequential manner. This allowed us to perform model fitting in time-efficient way, but has a risk of missing model parameter combinations which might not necessarily stem from the best-fit decision model yet provide a better fit to the  overall data. This limitation could be addressed in future by fitting the models to decision outcome, response times, and confidence data simultaneously.

\subsection{Implications}
Better understanding of human cognition can lead to better human-machine interaction~\cite{flemisch_2008, schurmann_personalizing_2020} and contribute to safe and efficient intelligent transportation systems~\cite{markkula_models_2018}.  We believe that our findings in particular can help improve driver assistance systems and automated vehicles which could utilize the insight into driver confidence for more personalized automation~\cite{yan_developing_2016, yan_building_2017}.

More fundamentally, our study exemplifies how real-world tasks can be translated into controlled experimental paradigms, providing a new kind of testbed for theories and models developed in basic science using distilled tasks~\cite{maselli_beyond_2023, boag_evidence_2023}. We believe such cross-pollination between fundamental cognitive science and real-world behavior will both enrich theoretical developments and enable new applications.

\subsection{Conclusions}
The study addressed the relation between confidence and the time-to-arrival and distance gap of the oncoming vehicle, the decision response time, and the action dynamics of the performed turn. We investigated to what extent confidence judgments can be captured by four simple cognitive models which have as a premise that confidence and decisions are based on the same evidence accumulation process. We conclude that:
\begin{itemize}
    \item time-to-arrival and distance gap to the oncoming vehicle affect the confidence judgments of left-turning drivers. In particular, confidence in “go” decisions is positively related to the time-to-arrival and the distance gap. Confidence in “wait” decisions relates negatively to the time-to-arrival and distance gap.

    \item confidence relates negatively to the response time regardless of the decision outcome.

    \item velocity profile during the turn and the distance to the centre of the intersection relate to the confidence of the participant. Participants who were more confident in their decision drove in general faster during the turn. For “go” decisions, low confidence judgments appeared to be associated with corner-cutting behavior.

    \item participants’ reported confidence in left-turn gap acceptance/rejection decisions can best be explained by an extended drift-diffusion model in which confidence judgment is based on post-decision evidence accumulation.  
\end{itemize}

\section*{Data availability}
The supplementary information, all raw and processed data, and the analysis and modelling scripts are available at \url{https://osf.io/tgexp}. \\

\bibliography{sn-bibliography}

\end{document}